\documentclass[aps,twocolumn,floats,nofootinbib,prl]{revtex4}
\usepackage{graphics,graphicx,epsfig}
\usepackage{amssymb,color}
\usepackage{epsf,epstopdf,wrapfig}
\usepackage {amsmath}
\usepackage{cancel}
 
\definecolor{green(ryb)}{rgb}{0.4, 0.69, 0.2}

\newcommand{\beq}{\begin{equation}}
\newcommand{\eeq}{\end{equation}}
\newcommand{\beqn}{\begin{eqnarray}}
\newcommand{\eeqn}{\end{eqnarray}}

\begin{document}

\title{How accurately must cells measure the concentrations of signaling molecules?}

\author{Marianne Bauer$^{a,b}$ and William Bialek,$^{b,c}$}

\affiliation{$^a$Department of Bionanoscience, Kavli Institute of Nanoscience Delft, Technische Universiteit Delft, Van der Maasweg 9, 2629 HZ Delft, the Netherlands\\
$^b$Joseph Henry Laboratories of Physics and Lewis--Sigler Institute for Integrative Genomics,   Princeton University, Princeton, NJ 08544 USA\\
$^c$Center for Studies in Physics and Biology, Rockefeller University, 1230 York Avenue, New York, NY 10065 USA} 

\date{\today}

\begin{abstract}
Information of relevance to the organism often is represented by the concentrations of particular molecules inside a cell.  As outside observers we can now measure these concentrations precisely, but the cell's own mechanisms must be noisier.  We show that, in certain limits, there is a universal tradeoff between the information capacity of these concentration measurements and the amount of relevant information that is captured, a version of the information bottleneck problem.  This  universal tradeoff  is confirmed, quantitatively, in an analysis of the positional information encoded by the ``gap genes'' in the developing fly embryo.
\end{abstract}

\maketitle

In the physics laboratory, and in engineered devices, we are used to information being represented by electrical or optical signals.  While the brain also uses electrical signaling, inside living cells information often is represented by the concentrations of particular molecules.   The absolute concentrations of these molecules, and even their total number, can be quite small.  As a result there has been considerable interest in understanding the physical limits to this molecular signaling \cite{berg+purcell_77,bialek+setayeshgar_05,bialek+setayeshgar_08,endres+wingreen_09,mora+wingreen_10,kaizu+al_14,mora+al_15,tenWoldereview,mora+nemenman_19}, the strategies that cells can use to maximize information in the face of these limits \cite{tkacik+al_08,tostevin+wolde_09,tkacik+walczak_11,siggia+vergassola_13,becker+al_15,wolde+al_16}, and  the implications  for cellular function.

One approach to understanding signaling via molecular concentrations is to explore increasingly realistic models of the microscopic events \cite{ackers+al_82,ptashne+gann_02,bintu+al_05,hnisz+al_17,morrison+al_21,zoller+al_22,MartinezCorral}.
Recently we suggested a different approach, in which we ask more abstractly about the implications of noise in the response, or more precisely about the limited information capacity of the cell's ``measurements'' of concentration \cite{bauer+al_21,bauer_22}.  To formalize the problem, we imagine that some relevant signal $x$ is represented by the concentrations of $K$ different molecules, which we write as ${\mathbf g} \equiv \{g_1 , \, g_2,\, \cdots ,\, g_K \}$.  The cell does not have access to the exact values of ${\mathbf g}$, but only to some variables that constitute intermediates in the response.  As an example, if the molecules act by binding to particular sites along the cell's DNA and  thereby regulating the expression of downstream genes, the intermediate variable might be the average occupancy of these binding sites over some relevant time window, or the state of the enhancers built out of groups of these binding sites \cite{ptashne+gann_02,bintu+al_05,furlong+levine_18}.  Independent of molecular details,  this intermediate variable can carry only a limited amount of information about the real concentrations; in this sense it is a ``compressed'' representation \cite{cover+thomas_91}, and we will refer to this  representation as $C$.

Different mechanisms inside the cell will generate different mappings ${\mathbf g} \rightarrow C$, and in general we expect this mapping to be noisy, so it will be described by a probability distribution $P(C|{\mathbf g})$.  What would be useful for the cell is to capture as much information as possible about the relevant variable $x$, subject to the constraint that the information about $\mathbf g$ is limited.  This means that the best mapping  ${\mathbf g} \rightarrow C$ is one that maximizes
\begin{equation}
{\cal U} = I(C; x) - \lambda I(C; {\mathbf g}) ,
\label{IBdef}
\end{equation}
where $\lambda$ is a Lagrange multiplier to implement the constraint on information about $\mathbf g$.  This is an example of the information bottleneck problem \cite{bialek+al_99}, which arises in contexts ranging from text classification \cite{slonim+tishby_00}  to the analysis of deep networks  \cite{tishby_zaslavsky_77,saxe,alemi} and neural coding  \cite{schneidman+al_03,buesing+maass_10}.

Optimizing $\cal U$ will define a bounding curve,  which shows the minimum $I(C; {\mathbf g})$ needed to reach a criterion level of $I(C; x)$.  If the cell's measurements of concentration become more precise, then $I(C; {\mathbf g})$ becomes larger, but since $C$ depends on $\mathbf g$ and not directly on $x$ we always have $I(C; x) \leq I({\mathbf g}; x)$.  The question is how close the cell can come to capturing all this available information given limits on the precision of its response.

The standard approach to solving the information bottleneck problem is to assume that $C$ is discrete, so that the mapping ${\mathbf g} \rightarrow C$ becomes a kind of clustering, and the optimal $P(C|{\mathbf g})$ obeys a self--consistent equation that can be solved iteratively.  There is a separate bounding curve for each choice of the cardinality $||C||$, and the full structure emerges as we let $||C|| \rightarrow\infty$.  We have applied this approach to the representation of positional information by the gap gene expression levels in the early fly embryo \cite{bauer+al_21}.  The results showed, for example, that the optimal mappings for individual genes are close to an intuitive thresholding model, but that optimal compression of multiple genes depends crucially on combinatorial interactions among these molecules, as seen experimentally, for example in the regulation of the pair rule genes by the gap genes.  But this discussion seemed to depend on  details of the  genetic network in the fly embryo, and missed the possibility that there is something more universal about the tradeoff between $I(C; {\mathbf g})$ and $I(C; x)$.  Our goal  is to uncover this (asymptotically) universal tradeoff, which emerges in an intuitive limiting regime.

In many cases the relation between $\mathbf g$ and $x$ is smooth if noisy.   
To be concrete, we write the mean concentrations at each $x$ as $\langle g_\mu\rangle_x$, which are smooth functions of $x$, and  fluctuations around these means are described by a covariance matrix $\langle \delta g_\mu \delta g_\nu \rangle_x$.  To first approximation, the probability distribution $P({\mathbf g}|x)$ can be seen as Gaussian, fully described by these conditional moments.  If $\langle g_\mu\rangle_x$ is smooth, it is reasonable to think that the compressed variable $C$ also will have a smooth relation to other variables.  Further, although a single measurement of concentration might be noisy, one can imagine that the noise level is smaller if we think about the encoding of the relevant variable $x$.  These observations suggest that we search for optima in which $C$ is a continuous variable, and that at appropriate points we can take a small noise limit.  Concretely, let us assume that
\begin{equation}
P(C|{\mathbf g}) = {1\over\sqrt{2\pi\sigma_C^2({\mathbf g})}} \exp\left[ - {{(C - \bar C ({\mathbf g}))^2}\over {2\sigma_C^2 ({\mathbf g})}}\right] ,
\label{gauss1}
\end{equation}
so that optimization of $\cal U$ now corresponds to finding the optimal functions $\bar C ({\mathbf g})$ and $\sigma_C ({\mathbf g})$.  

With the Gaussian approximation of Eq (\ref{gauss1}) we can immediately write \cite{cover+thomas_91,bialek_12}
\begin{equation}
I(C;{\mathbf g}) = S(C) - {1\over 2}{\bigg\langle} \log_2\left[2\pi e \sigma_C^2 ({\mathbf g})\right]{\bigg\rangle},
\end{equation}
where $S(C)$ is the entropy of the variable $C$.  Importantly this entropy is finite in the limit of small noise, since $C$ must be tied, even if implicitly, to the relevant variable $x$.  In particular, if the effective  noise in estimating $x$ from $\mathbf g$ is small compared with the scale on which the distribution  $P(x)$ varies, then this relationship becomes nearly deterministic \cite{tkacik+al_08}, and we can write 
\begin{eqnarray}
P(C) &=& P(x) {\bigg |}{{d\bar C}\over{dx}}{\bigg |}^{-1}\\
S(C) &=& S(x) + {\bigg\langle} \log_2 {\bigg |}{{d\bar C}\over{dx}}{\bigg |} {\bigg\rangle} ,
\end{eqnarray}
where the dependence of $\bar C$ on $x$ is  through $\mathbf g$,
\begin{equation}
{{d\bar C}\over{dx}} = \sum_\mu {{d\langle g_\mu\rangle_x}\over{dx}} {{\partial \bar C}\over{\partial g_\mu}}{\bigg |}_{{\mathbf g} = \langle {\mathbf g}\rangle_x}  ,
\end{equation}
again working in a small noise limit.

To compute the information which $C$ conveys about $x$ we need the distribution
\begin{equation}
P(C|x) = \int d {\mathbf g} P(C|{\mathbf g}) P({\mathbf g}|x) ,
\end{equation}
which in general is complicated,  but if noise is small we can again  make a Gaussian approximation.  The variance of $C$ at fixed $x$  has two components, one from the variance at fixed $\mathbf g$, and one from the (co)variance of $\mathbf g$ at fixed $x$, 
$\langle \delta g_\mu \delta g_\nu \rangle_x$:
\begin{eqnarray}
\langle (\delta C)^2 \rangle_x &=& \sigma_C^2 ({\mathbf g}){\bigg |}_{{\mathbf g} = \langle {\mathbf g}\rangle_x} + A\\
A &=&  \sum_{\mu\nu} {{\partial \bar C}\over{\partial g_\mu}}  \langle \delta g_\mu \delta g_\nu \rangle_x {{\partial \bar C}\over{\partial g_\nu}} {\bigg |}_{{\mathbf g} = \langle {\mathbf g}\rangle_x} ,
\label{def_A}
\end{eqnarray}
With this we have
\begin{equation}
I(C; x) = S(C) - {1\over 2}{\bigg\langle} \log_2\left[2\pi e \langle (\delta C)^2 \rangle_x \right]{\bigg\rangle} ,
\end{equation}
and all the ingredients needed to express the objective function $\cal U$.

We notice that $\cal U$ is a local functional of $\bar C ({\mathbf g})$ and $\sigma_C^2 ({\mathbf g})$, so we can use the calculus of variations in a familiar way.  Optimizing with respect to $\sigma_C^2 ({\mathbf g})$ is especially straightforward, and we find
\begin{eqnarray}
{{\partial {\cal U}}\over{\partial \sigma_C^2 ({\mathbf g})}} &=& 0  \\
\Rightarrow \sigma_C^2 ({\mathbf g}) &=& {\lambda \over{1-\lambda}} A,
\end{eqnarray}
with $A$ from Eq (\ref{def_A}).  This makes sense, since it tells us that the precision of encoding $\mathbf g$ in $C$ should be related to the scale of the fluctuations in $\mathbf g$ when the relevant variable $x$ is fixed.  At this optimum we have
\begin{eqnarray}
I(C; {\mathbf g}) &=&  S(x) - {1\over 2}{\bigg\langle} \log_2\left[{{2\pi e}\over B} {\lambda\over{1-\lambda}} \right]{\bigg\rangle}
\label{R1A}\\
I(C; x) &=& S(x) -  {1\over 2}{\bigg\langle} \log_2\left[{{2\pi e}\over B} {1\over{1-\lambda}} \right]{\bigg\rangle},
\label{R1B}
\end{eqnarray}
with $B = (1/A) [{{d\bar C}/{dx}}]^2$.  Optimizing $\cal U$ now is equivalent to maximizing $B$ with respect to the function $\bar C ({\mathbf g})$.

Notice that if there is only one variable $g$, then
\begin{eqnarray}
A &=& A_1 = {\bigg |}{{d\bar C}\over{dg}}{\bigg |}^2  \langle (\delta g)^2\rangle\\
B &=& B_1 = {1\over {A_1}} {\bigg |}{{d\bar C}\over{dx}}{\bigg |}^2  .
\end{eqnarray}
Since are working in the limit where noise is small, 
\begin{eqnarray}
{{d\bar C}\over{dx}} &=& {{d\bar C}\over{dg}} \cdot {{d\langle g\rangle_x}\over {dx}} \\
\Rightarrow B &=& {1\over{ \langle (\delta g)^2\rangle}}  {\bigg |}{{d\langle g\rangle_x}\over {dx}}{\bigg |}^2 ={1\over {\sigma_x^2 ({g})}} ,
\end{eqnarray}
where in the last step we recognize $\sigma_x$ as the error bar in estimating $x$ from the vector of concentration $g$ \cite{dubuis+al_13,tkacik+al_15}; this identification itself is correct only in the low noise limit.  In this case $B$ is determined independent of our choice for ${\bar C} ({\mathbf g})$, and we can go further to show that the expectation values which appear in Eq (\ref{R1A}, \ref{R1B}) are related to the available information $I(g; x)$. Perhaps surprisingly, we will see that with multiple variables we arrive at the same answer by optimizing the function ${\bar C} ({\mathbf g})$.

Because $\cal U$ depends on the derivatives of $\bar C$ with respect to the components of the vector $\mathbf g$, the Euler--Lagrange equation $\delta {\cal U}/\delta \bar C = 0$ corresponds as usual to the divergence of some vector $\mathbf V$ vanishing.  But if the derivatives $d\langle g_\mu \rangle_x /dx$ vanish at extreme values of the relevant variable $x$, then $\mathbf V$ itself must vanish.  The result is that we can write an explicit, self--consistent expression for the gradient of $\bar C$,
\begin{eqnarray} \label{eq:optim_c}
{{\partial\bar C ({\mathbf g})}\over{\partial  g_\mu}} &=&
D \sum_\nu \left[\left( \langle \delta g \delta g\rangle_x \right)^{-1} \right]_{\mu\nu}  {{d \langle g_\nu \rangle_x}\over{dx}} \\
D &=&  {{\nabla_g\bar C \cdot \langle \delta g \delta g \rangle_x \cdot \nabla_g\bar C}
\over{\nabla_g\bar C \cdot \nabla_x \langle g \rangle}}  .
\end{eqnarray}
This is complicated, but yields a simple expression for $B$,
\begin{eqnarray}
B &=&  \sum_{\mu\nu} {{d \langle g_\mu \rangle}\over{dx}} \left[\left( \langle \delta g \delta g\rangle_x \right)^{-1}\right]_{\mu\nu} {{d \langle g_\nu \rangle}\over{dx}}\\
&=& {1\over {\sigma_x^2 ({\mathbf g})}} ,
\end{eqnarray}
where  $\sigma_x ({\mathbf g})$ now is the error bar in estimating $x$ from the entire vector of concentrations $\mathbf g$ \cite{dubuis+al_13,tkacik+al_15}. 
Then we can identify the information that the concentrations $\mathbf g$ provide about $x$,
\begin{equation}
I({\mathbf g}; x) = S(x) - {1\over 2}{\bigg\langle} \log_2\left[2\pi e \sigma_x^2 ({\mathbf g}) \right]{\bigg\rangle} .
\end{equation}

\begin{figure*}[th!]
\centerline{\includegraphics[width = \linewidth]{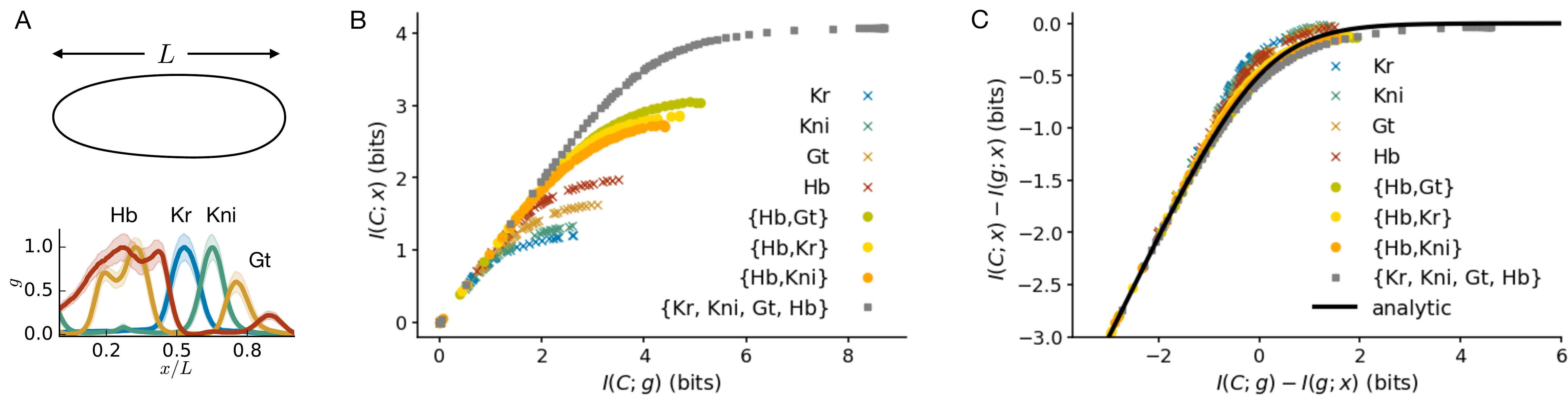}}
	\caption{   (A) The four gap genes Hb, Kr, Kni and Gt are expressed in varying concentrations $g$ along the anterior--posterior axis of the fly embryo $x$ \cite{petkova+al_19}; $L\sim 0.5\,{\rm mm}$. (B) The information plane $I(C;x)$ vs $I(C;g)$ for optimized $C$ when $g$ is each of the four gap gene expressions separately, some exemplary pairs, and all gap gene expressions together; computations follow the methods of Ref \cite{bauer+al_21}. (C) Collapse of the data as predicted from Eqs (\ref{IB1}) and (\ref{IB2}), with the boundary curve from the universal tradeoff, Eqs (\ref{IB1}) and (\ref{IB2}), in black.
\label{fig_test}}
\end{figure*}

Finally, putting the different terms together we find that along the bounding curve
\begin{eqnarray}
I(C; x) &=& I({\mathbf g}; x) -  {1\over 2}  \log_2\left( {1\over{1-\lambda}} \right) \label{IB1}\\
I(C; {\mathbf g}) &=&  I({\mathbf g}; x) - {1\over 2}  \log_2\left( {\lambda\over{1-\lambda}} \right) .\label{IB2}
\end{eqnarray}
We see that with $0 < \lambda < 1$, the information $I(C;x)$ that is captured about the relevant variable is less than the available information $I({\mathbf g}; x)$, as it must be.   As $\lambda\rightarrow 0$ this gap closes, but at the expense of requiring an increasing information capacity in the mapping ${\mathbf g} \rightarrow C$.  Taken together, Eqs (\ref{IB1}) and (\ref{IB2}) define the bounding curve in the information plane, as shown in Fig \ref{fig_test}.

The result in Eqs (\ref{IB1}) and (\ref{IB2}) is surprising, because all details of the underlying system have disappeared.  This  asymptotically universal bounding curve suggests there is a tradeoff between the capacity of the cell to measure the concentration of signaling molecules and the resulting ability to capture relevant information, independent of molecular mechanisms, at least in some regime. 
Are real cells in this regime?

The gap genes form a network that is crucial to the early events of fly development \cite{nusslein-vollhard+wieschaus_80,jaeger_11}.  These four genes take inputs from primary maternal morphogens and in turn drive the striped patterns of pair--rule gene expression.  The local concentrations of gap gene proteins provide enough information to specify position to $\sim 1\%$ accuracy along the anterior--posterior axis, and this is the precision with which the stripes are positioned \cite{dubuis+al_13,Liu2013,antonetti+al_18}. The algorithm that achieves optimal readout of this positional information predicts, quantitatively, the distortions of the pair--rule stripes in mutant flies where individual maternal inputs are deleted \cite{petkova+al_19}.

The concentrations of the gap gene proteins ($\mathbf g$) encode information about position ($x$), providing an example of the problem we have been discussing;   Fig \ref{fig_test}A shows the means and standard deviations of $\mathbf g$ as a function of $x$.  With four gap genes, we can analyze the information that they carry individually, or in groups; Figure \ref{fig_test}B shows the numerical solutions to the information bottleneck problem---minimizing $\cal U$ in Eq (\ref{IBdef})---for individual genes, exemplary combinations of two genes, and the full set of four genes together \cite{bauer+al_21,bauer_22}.  Figure \ref{fig_test}C shows that these curves collapse when shifted by the mutual information $I({\mathbf g}; x)$, as predicted in Eqs (\ref{IB1}, \ref{IB2}).  Notice that these shifts vary by up to $\sim 2.5\,{\rm bits}$ across the different groups.   We see that these real examples follow the predictions of the universal tradeoff  (in black) quite accurately. The case with all four genes is closest to the theory, which makes sense since this is the case where our small noise approximation is most accurate.

The fact that a real system follows the universal tradeoff in Eqs (\ref{IB1}) and (\ref{IB2}) invites us to consider the implications.  We notice that the theoretical prediction is close to being a line of unit slope, $I(C;x) \sim I(C;{\mathbf g})$, ending in saturation at  $I(C; x ) = I({\mathbf g}; x)$.   We can never have $I(C;x) > I(C;{\mathbf g})$, and it is interesting that, under fairly general conditions, it is possible to approach this maximal efficiency: if noisy mechanisms can keep only $I$ bits of information about the signaling molecule concentrations, then it is possible for all of these $I$ bits to be relevant for the organism, even when the mappings among signals are complicated, as in the pattern of gap gene expression vs position;  this is not true in general  \cite{NgampruetikornSchwab}.  We can see how close this tradeoff is to the diagonal $I(C;x) =I(C;{\mathbf g})$ by calculating $I(C; x)$ at the point $I(C;{\mathbf g}) = I({\mathbf g},x)$; from Eq (\ref{IB2})   this happens at $\lambda = 0.5$, where Eq (\ref{IB1}) predicts that  $I(C;x) = I({\mathbf g},x) - 0.5 \,{\rm bits}$. Thus, in the regime we are considering, sensors with ``just enough'' capacity to transmit all the information provided by the signaling molecules can come close to deploying all this capacity for the relevant information.  

On the other hand, it is worth emphasizing that no real mechanism can extract {\em all} of the information that is available about $x$ from a perfect measurement of $\mathbf g$.  If the available information is $I$, then to get within $\epsilon\,{\rm bits}$ Eq (\ref{IB1}) tell us that we need a mechanism with capacity
$I(C;{\mathbf g}) \sim I - (1/2)\log_2 (2\epsilon\ln 2)$ at small $\epsilon$.  If the cell needs to make a binary decision, then making errors with small probability $q$ is  equivalent to losing $\epsilon \sim q\log_2(1/eq)\,{\rm bits}$.  This means, for example, that if the initial signals $\mathbf g$ are just sufficient to provide one bit of information, the cell would need to read these signals with $\sim 2.5\,{\rm bits}$ of accuracy in order to keep errors below $\sim 1\%$.  The requirements are even more stringent if we imagine that the initial signal $\mathbf g$ is processed through several layers.  The perhaps surprising conclusion is that mechanisms with one bit of information capacity are not sufficient for cells to make reliable binary decisions.

\begin{acknowledgments}
We thank our experimental colleagues T Gregor, MD Petkova, and EF Wieschaus, whose results inspired these ideas.  This work was supported in part by the National Science Foundation through the Center for the Physics of Biological Function (PHY--1734030), by fellowships from the Simons Foundation and  the John Simon Guggenheim Memorial Foundation (WB), and by a start-up grant from the Bionanoscience Department at TU Delft (MB).
\end{acknowledgments}

\end{document}